\documentclass[referee]{aa} 
\usepackage{graphicx}

\newcounter{sub}
\newcounter{subeqn}[sub]
\newcommand\be{\begin{equation}}
\newcommand\ee{\end{equation}}
\newcommand\lp{\left(}
\newcommand\rp{\right)}

\newcommand\st{\stepcounter{sub}}

\newcommand\bea{\begin{eqnarray}}
\newcommand\eea{\end{eqnarray}}
\newcommand\bean{\begin{eqnarray*}}
\newcommand\eean{\end{eqnarray*}}

\newcommand\oomega{\mbox{\boldmath $\omega$}}

\newcommand\unit{\mbox{\boldmath $e$}}

\newcommand\m{{\bf m}}

\newcommand\LL{{\bf L}}

\begin{document}

   \thesaurus{01     
              (
             )}
%

  \title{Radiation pattern of the isolated pulsar PSR B1828-11}
  
   \author{Vahid Rezania }

   \offprints{V. Rezania \\ email: vrezania@phys.ualberta.ca\\
                   $^*$ Present address}

    \institute{Department of Physics, University of Alberta,
              Edmonton AB, Canada T6G 2J1 $^*$ \\       
              Institute for Advanced Studies in Basic Sciences,
               Gava Zang, Zanjan 45195, Iran
                    }

   \date{Received / accepted }

\maketitle

\abstract{
Based on the free precession model of the isolated pulsar PSR B1828-11,
Link \& Epstein (\cite{LE01}) showed that the observed pulse durations
require the radio beam to have a non-standard shape: 
the beam duration is larger for beam sweeps farthest from the dipole axis.
In their analysis they assumed that the actual precession period is $\simeq
500$ d.  Recent theoretical studies suggested that the actual
precession period might be $\simeq 1000$ d as seen in observations 
(Rezania \cite{Rez02}, Wasserman \cite{Was02}).
In this paper, in a good agreement
with the observed data (Stairs et al. \cite{SLS00}), 
we model the changes of the pulse shape in a precession cycle with period
$\simeq 1000$ d and find that the variation of the pulse duration follows 
from a {\it standard} beam pattern in each cycle. 

}\\

\section{Introduction}
Analysis of the long-term observation of the spin behavior of the
isolated pulsar PSR B1828-11 reveals periodic variations both in
the pulse shape and the slow-down rate of the pulsar and shows strong
Fourier power
at periods of~$\simeq 1000, 500, 250$, and $167$ d
(Stairs et al. \cite{SLS00}).
Close correlations of the periodic changes in the pulse shape and
duration with
variations in the spin-down rate of the pulsar strongly 
suggest that the star's spin axis is freely precessing around
the star's symmetry axis.

Recently Link \& Epstein (\cite{LE01}) studied the behavior
of the observed pulse duration of PSR B1828-11 based on the 
free precession model.  Since both $500$ d and $250$ d Fourier components
have dominant contributions in the observed variations
of period residual $\Delta p$, its derivative $\Delta \dot p$, and
pulse shape of the beam, they suggested that the actual
free precession period of the star is close to the strongest
Fourier component $\simeq 500$ d.   Then a coupling of nearly orthogonal
magnetic
dipole moment to the star's spin axis would provide the observed harmonic
at period $250$ d.  On this basis, they modeled the pulse duration
variations as a function of beam's sweep angle $\Delta\Theta$
(see below for definition),
and found that the beam pattern of radiations {\it must} be non-standard:
the beam duration is larger for beam sweeps farthest from the dipole axis.  
As a result, both upper and lower parts of the emission region are
wider than its middle, see Figure 2.
As they mentioned this is {\it required} by the precession interpretation
of PSR B1828-11.

Recent theoretical investigations on the free precessing motion of
PSR B1828-11 provide new explanations for the reported data with the
fundamental period $\sim 1000$ d, which was originally suggested by
Jones \& Andersson (\cite{JA01}).  Rezania (\cite{Rez02}) considered the
case in which the 
magnetic field of the star varies with time while the star is precessing.
Then he found a condition under which 
a coupling of the star's crust with the time-varying magnetic radiation
torque would produce the whole 
observed Fourier spectrum consistently.
In a good agreement with the data he found that the fundamental
precession period would be $\simeq 1000$ d.   
Alternatively, Wasserman (\cite{Was02})
showed that in general an oblique rotator must precess.   By analysis of
the mechanical energy of the system,
he found that the minimum energy state for such star is a state where
the star precesses.  For strong magnetic stresses
in the star's type II superconductor core 
he estimated the precession period as
$P_{\rm pre}\simeq 2460 {\rm\; d}/(\beta\cos\chi B_{12}H_{15})$, where
$B_{12}=B/10^{12}$ G is the star's magnetic field strength,
$H_{15}=H/10^{15}$ G corresponds to the first critical field strength
in a type II superconductor ($H_{cr}\sim 10^{15}$ G), $\beta\sim 1$
and $\chi$ is the inclination angle of magnetic symmetry axis.  For the case
PSR B1828-11 with $B\sim 5\times 10^{12}$ G, 
the precession period will be
$P_{\rm pre}\sim 1000$ d, if the inclination angle is
$\chi\simeq 60^\circ$.

In this paper we study the pulse shape of PSR B1828-11 by assuming that
the actual precession period of the star is $\simeq 1000$ d rather than
$\simeq 500$ d. We note that
Link \& Epstein's calculations give the right behavior of the pulse
shape variations {\it provided} the actual precession period is close to
$\simeq 500$ d {\it only}.  For the case $P_{\rm pre}\simeq 1000$ d, their
analysis gives an incorrect prediction for the observed shape variations.  
Here, we generalize link \& Epstein's analysis and show that,
for $P_{\rm pre}\simeq 1000$ d, to get the
correct behavior for the shape variations (compared with data)
the radio beam of PSR B1828-11 must have the {\it standard} pattern. 


\section{The model}
Let $x$, $y$, and $z$ represent an inertial coordinate system $\cal S$, 
and an observer is in the $x-z$ plane with $x>0$ and $z>0$.
Now consider a rigid, biaxial rotating star with angular velocity $\oomega$, 
the principal axes $\unit_1, \unit_2, \unit_3$, and corresponding principal 
moment of inertia $I_1=I_2\neq I_3$.  
The star's angular momentum $\LL$, along the $z$-axis in $\cal S$, is
misaligned to the star's symmetry axis $\unit_3$ by a wobble
angle $\theta$, ie. $\LL\cdot \unit_3=L\cos\theta$, see Figure 1.   
To study the pulse shape variations we look at 
the variation of the polar angle of the beam with
respect to $\LL$ (fixed in $\cal S$) that is equal to the polar
angle of the magnetic dipole moment $\m$.\footnote{
Note that though the pulsar beam's direction is not necessarily
in the same direction as the dipole moment $\m$, for simplicity 
Link \& Epstein (\cite{LE01}) defined a pulse as occurring when the
azimuthal angle $\Phi$ of the magnetic dipole equals to the azimuth of the
observer.     
As a result, $\dot\Phi$ is the
observed pulse frequency.  To compare our concluding result with one
obtained by Link \& Epstein (\cite{LE01}), we use the same definitions.}
The azimuthal and polar angles of
the magnetic dipole moment $\m$, $\Phi$ and $\Theta$ in the inertial frame
$\cal S$ are given by $\tan\Phi=m_y/m_x$ and 
$\cos\Theta=m_z/m=\sin\theta\sin\psi\sin\chi+\cos\theta\cos\chi$, 
where $\theta$ is the wobble angle,
$\psi=\tan^{-1}\omega_1/\omega_2
=\pi/2-\omega_pt - \beta$ ($\omega_i=\unit_{i}\cdot\oomega$)
and $\chi$ is the inclination of magnetic field
symmetry axis from the star's symmetry axis $\unit_3$, ie.
$\m\cdot\unit_{3}=m\cos\chi$.  Here
$\omega_p=2\pi/P_{\rm pre}$ is free precession frequency and $\beta$ is a
constant phase.
Expanding $\cos\Theta$ to the first order in $\theta$ one finds
$\Theta=\chi-\theta\cos(\omega_pt+\beta)$. 
The latter shows the beam polar angle $\Theta$ changes
sinusoidally about $\chi$, as the star precesses.

Now let $\gamma\equiv\xi+\chi$ be the polar angle of the observer
in $\cal S$
(the constant angle between the observer and the angular momentum vector),
see Figure 1.  Following Link \& Epstein (\cite{LE01}) we define the
sweep angle $\Delta \Theta$ as the difference in polar angle of
observer, $\gamma$, and the dipole, $\Theta$ at the time of the pulse:
\st
\be\label{DTheta} 
\Delta \Theta\equiv\gamma-\Theta =
\xi+\theta\cos(\omega_pt+\beta)+{\cal O}(\theta^2).
\ee
Here $\xi$ is a free parameter which will be fixed later by fitting
the data.
Link \& Epstein assumed
that the pulse duration $w$ is a function of $\Delta \Theta$ only, and has 
an extremum at $\Delta \Theta=0$.
Then they expanded the pulse duration $w$ in terms of $\Delta\Theta$
as $w=w_0+w_2(\Delta\Theta)^2$.  
As it is clear, the latter expression would provide both $P_{\rm pre}$
and $P_{\rm pre}/2$ components (due to the $\cos(\omega_pt+\beta)$
and $\cos^2(\omega_pt+\beta)$ terms) 
in the pulse duration.  As a result, it gives the correct behavior of
the observed pulse duration for $P_{\rm pre}\simeq 500$ d.
But with $P_{\rm pre}\simeq 1000$ d, the $250$ d Fourier component is
missing.  
To get the correct behavior for
the pulse shape variations during a $1000$ d precession cycle, 
we take a more general expansion rather than Link \& Epstein (\cite{LE01})
as
\st\be\label{w}
w=w_0+w_2(\Delta\Theta)^2+w_4(\Delta\Theta)^4.
\ee
It is interesting to note that equation (\ref{w}) provides
the contribution of $P_{\rm pre}$, $P_{\rm pre}/2$,
$P_{\rm pre}/3$, and $P_{\rm pre}/4$ terms in the pulse duration.
So with $P_{\rm pre}\simeq 1000$ d, one would expect to observe $500$ d,
$333$ d, and $250$ d Fourier components in the pulse shape parameter.
As reported by Stairs et al. (\cite{SLS00}), the $1000$ d and $333$ d
components though small (in comparing with $500$ d and $250$ d components)
they have {\it non-zero} amplitude in the observed pulse shape parameter.
These terms are missing in the Link \& Epstein's model.

Without loss of generality we assume that $w_4=1$.
Following Stairs et al. (\cite{SLS00}), we define the shape parameter as 
$S=\frac{A_N}{A_N+A_W}$ where $A_N$ and $A_W$ are the fitted heights of
the narrower and wider standard profiles respectively, so that $S\simeq 0$
for the wide pulses and $S\simeq 1$ for narrow ones.  
As a result, one can relate the shape parameter $S$ to the pulse
duration $w$ of the observed beam by
\st\be\label{S}
w= {\rm max}(w)(1-S)+{\rm min }(w)S, 
\ee
where max$(w)$ and min$(w)$ are the maximum and minimum values of the beam 
duration in the precession cycle.
Combining equations (\ref{w}) and (\ref{S}), one finds for $w_2>0$
\st\be
S=1-\frac{w_2 + (\Delta\Theta)^2}{ w_2 + (|\xi|+\theta)^2}  \lp 
{\Delta\Theta\over\vert\xi\vert+\theta}\rp^2,
\ee
while for $w_2<0$ we have 
\st\be
S/S_0  = \left\{ \begin{array}{ll} 
\frac{|w_2| - (\Delta\Theta)^2}{|w_2| - (|\xi|+\theta)^2}    
 \lp {\Delta\Theta\over\vert\xi\vert+\theta} \rp^2 & 
\mbox{{\rm if}~~~$|w_2|>(|\xi|+\theta)^2 $}, \cr
1 - \frac{|w_2| - (\Delta\Theta)^2 }{|w_2| - (|\xi|+\theta)^2}  
\lp {\Delta\Theta\over\vert\xi\vert+\theta} \rp^2 & 
\mbox{{\rm if}~~~$|w_2|<(|\xi|+\theta)^2$}, 
\end{array}
\right.
\ee
where $S_0$ is a normalization factor.
It is clear that the shape parameter
depends on $\omega_p$, $\beta$, $\theta$, $\xi$, and both the sign and 
magnitude of $w_2$.  With the given values of $\theta\simeq 3^\circ$
and $P_{\rm pre}\simeq 1000$ d
the shape parameter is determined by $\xi$, $w_2$, and sign$(w_2)$ 
completely.   In a good agreement with data, we find that $w_2$ must be 
{\it negative} with magnitude larger but close to $|w_2|\geq 
(|\xi|+\theta)^2$.   Inserting the latter results in equation (\ref{w}), 
one can easily show that the pulse duration $w$ is bigger
when the beam sweeps closer to the dipole axis  
($\Delta\Theta=0$) rather than when it sweeps farther
($\Delta\Theta\neq 0$).  This means 
the beam pattern is {\it standard}:
the beam duration is smaller for beam sweeps farthest from the dipole axis.  
Our result is completely in contrast with the result of
Link \& Epstein (\cite{LE01}) who found that a
non-standard beam pattern
(the beam duration is larger for beam sweeps farthest from the dipole axis)  
is {\it required} to explain the data.  
In Figure 2 we sample the pulse profile for different observer's
viewing angles (the solid closed curve) and compare with the one (the dashed
parabola) obtained by Link \& Epstein (\cite{LE01}).  We find the
radiation pattern of PSR B1828-11 is very close to multipole field radiation
pattern with $(\ell, m)=(1, \pm 1)$ where $\ell$ and $m$ are orbital and
azimuthal numbers. 
As the star precesses,
the observer sees different sweeps corresponded to different viewing angles.
Line B corresponds to the viewing angle (closest to the dipole axis)
for which pulse is the widest, $S\simeq 0$ (narrowest, $S\simeq 1$)
provided the
star precesses with precession period $P_{\rm pre}\simeq 1000$ d
($P_{\rm pre}\simeq 500$ d).   Accordingly, the pulse profile
is narrower, $S\simeq 0.4$ (wider, $S\simeq 0.4$) for line A
and narrowest, $S\simeq 1$ (widest, $S\simeq 0$) for line C.

By choosing an appropriate values for $|w_2|$ and $\xi$ one can
easily fit the data.
We found an acceptable fit on the pulse shape data of PSR B1828-11 
by choosing $|w_2|\simeq 0.0034$ and $\xi=-0^\circ.01$
for the given values of $\theta= 3^\circ .2$ and $\omega_p= 2
\pi/1016$ d$^{-1}$, see Figure 3.  Here the value of
$(|\xi|+\theta)^2\simeq 0.0031$.   We note that our fit curve which is
calculated for free precession period $P_{\rm pre}=1016$ d, is
indistinguishable from the one that was calculated by Link \& Epstein
(\cite{LE01}) with free precession period $P_{\rm pre}=511$ d. 


\section{Summary}
Previous study on the pulse shape variations of the isolated
pulsar PSR 1828-11 by Link \& Epstein (\cite{LE01}) is based on the
free precessing star with precession period $\sim 500$ d.  
Recent theoretical studies (Rezania \cite{Rez02}, Wasserman \cite{Was02})
on the free precession of PSR B1828-11 suggested 
that the actual precession period might be $\simeq 1000$ d as seen in
observations.  
In this paper, by assuming that
the actual free precession period of the isolated pulsar PSR B1828-11 is
$P_{\rm pre}\simeq 1000$ d we studied the star's pulse shape variations
during a precession cycle.
To get the correct behavior in the pulse shape
variations in comparing with the observed data, we expanded the pulse
duration $w$ to forth order of $\Delta\Theta$ (the difference between polar
angles of the observer and dipole axis at the time of the pulse).  
The forth order term is not considered by Link \& Epstein (\cite{LE01}). 
We found that the pulse duration $w$ is larger 
when the beam sweeps closer to the dipole axis  
($\Delta\Theta=0$) rather than when it sweeps farther to the dipole axis 
($\Delta\Theta\neq 0$), see Figure 2.  This means 
that the variation of the pulse
duration follows from a {\it standard} beam pattern in each cycle. 
Our result is in contrast with the result of Link \& Epstein (\cite{LE01})
who found that a non-standard beam pattern is required to explain the
data.                                          

Further, our model provides the contribution other Fourier harmonics,
$1000$ d and $333$ d, in the pulse shape parameter as seen in data.
Although these terms have smaller (but non-zero) amplitudes rather
than $500$ d and $250$ d Fourier components, they are forbidden in the
Link \& Epstein's model.   In addition, by taking into account
a term $(\Delta\Theta)^6$ in equation (\ref{w}), one can explain 
the $167$ d Fourier component seen in data of PSR B1828-11 
(Stairs et al. \cite{SLS00}). The latter which corresponds
to $P_{\rm pre}/6$ Fourier component for $P_{\rm pre}\simeq 1000$ d
is also forbidden for $P_{\rm pre}\simeq 500$ d free precession model.

Finally, we note that new observations of the pulse shape variations of 
PSR B1828-11 would be able to determine its radiation pattern.
As predicted by Link \& Epstein (\cite{LE01}), in a $511$ d precession
cycle the observer would see both the broad upper and lower parts of
the radio beam.  This would be a critical observational evidence for the
proposed theoretical models, since by assuming
$P_{\rm pre}\simeq 1000$ d as the fundamental precession period, 
the middle of the beam will be broader.

\acknowledgements
I would like to thank S. M. Morsink and S. Sengupta for
their careful reading
the manuscript and useful discussions.
The author is also grateful to I. Stairs for useful discussions and
providing
the timing data for PSR B1828-11. I would like to thank Roy Maartens 
for continuing encouragement.  This research was supported by the Natural 
Sciences and Engineering Research Council of Canada.

\newpage
\begin{figure*}
\centering
\includegraphics{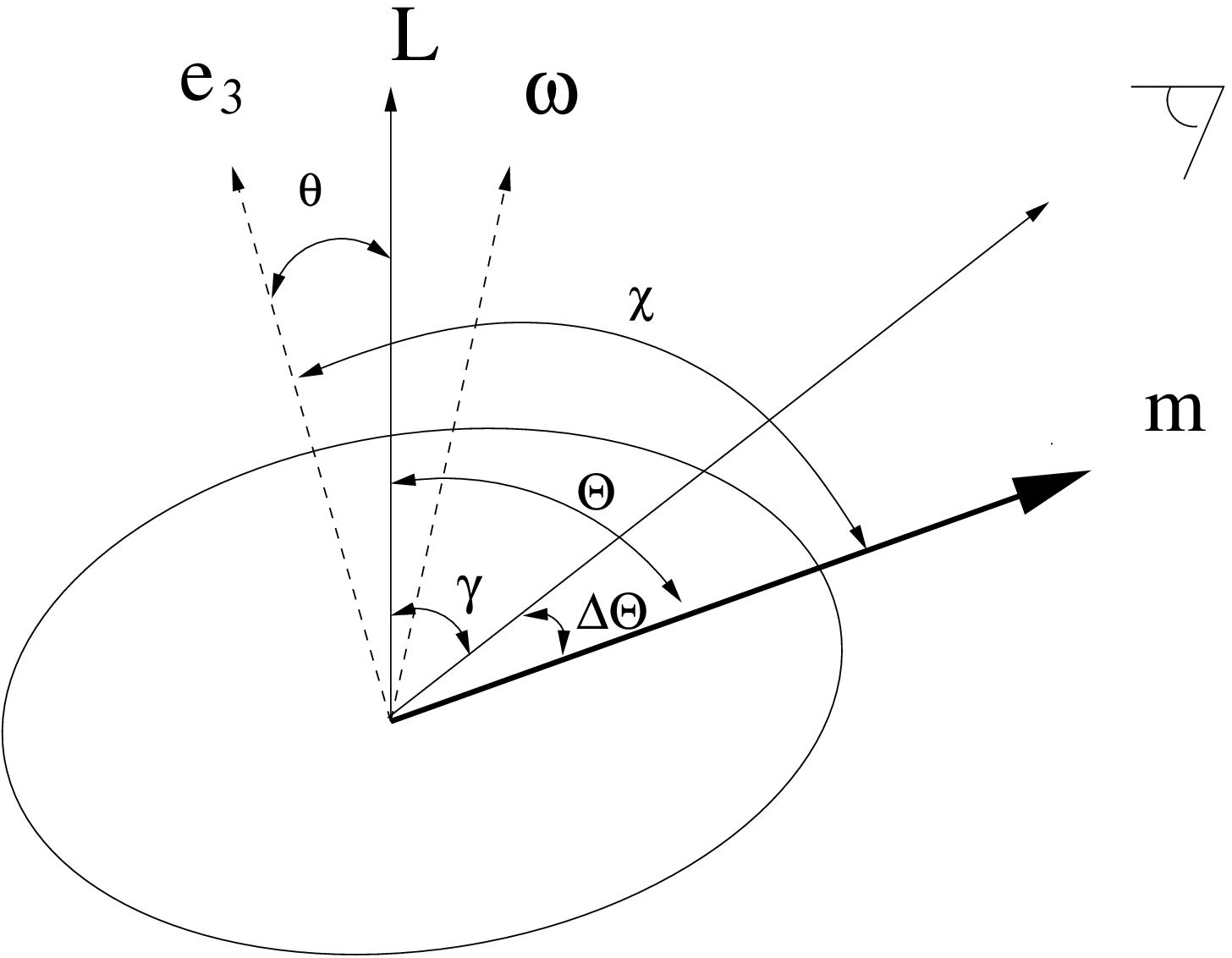}
\caption{ Observing geometry.  The angles defined at the instant the 
dipole moment $\m$ is in the plane containing the angular momentum $\LL$ and 
the observer.}
\label{Fig1}%
\end{figure*}
\newpage
\begin{figure*}
\centering
\includegraphics{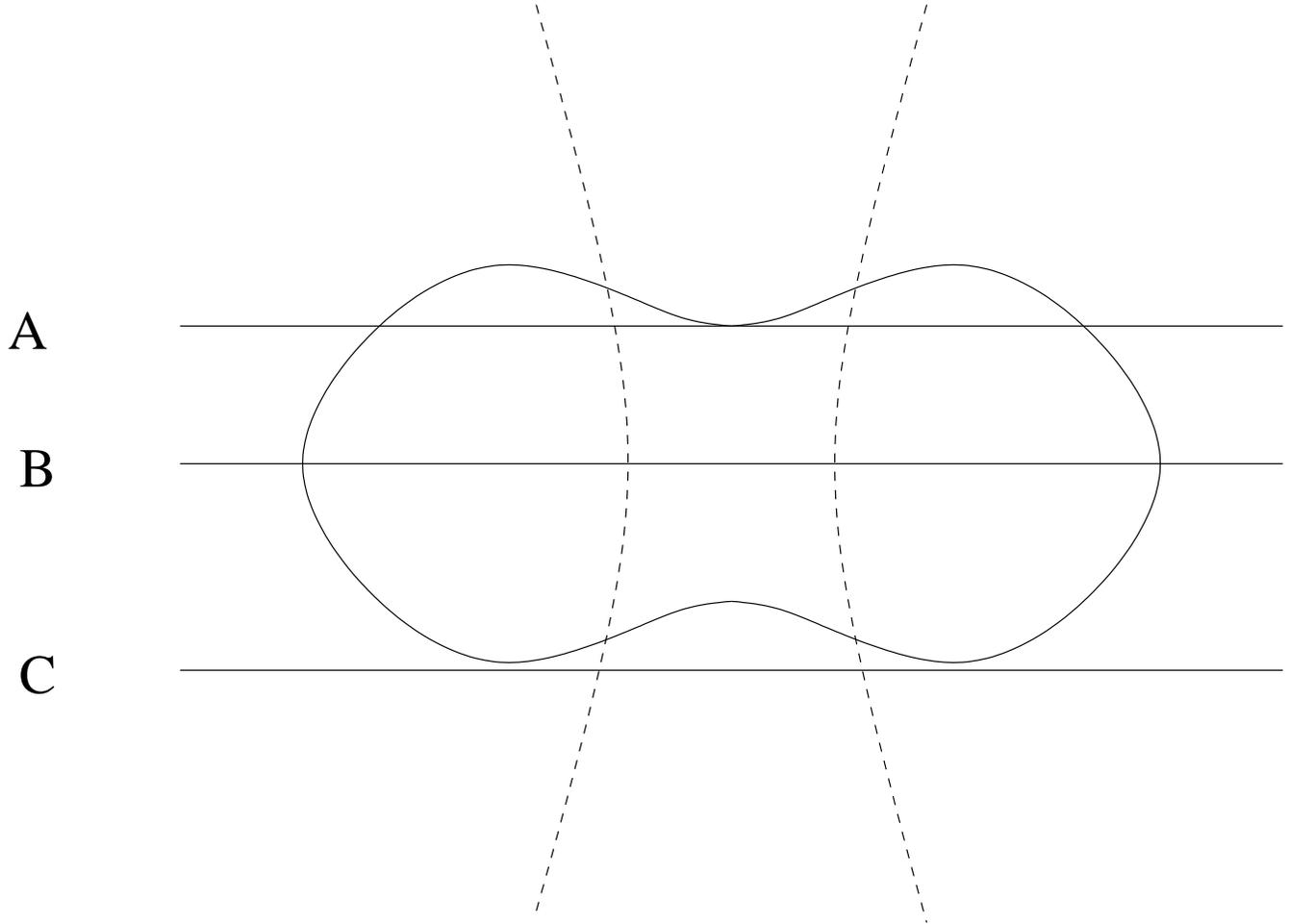}
\caption{ 
The beam pattern for the radio emission from PSR B1828-11.  The dashed
parabolas represent the beam pattern proposed by Link \& Epstein
(\cite{LE01}) with $P_{\rm pre}\simeq 500$ d, while the solid closed curve
correspond to the $1000$ d precession period.  The radiation
pattern is very close to multipole radiation pattern with
$(\ell, m)=(1, \pm 1)$.  The lines B, A, C represent the viewing
angle at which the beam is
the narrowest ($S\simeq 1$), wider ($S\simeq 0.4$),
and widest ($S\simeq 0$), respectively,
for Link \& Epstein's model, while correspond to
the widest ($S\simeq 0$), narrower ($S\simeq 0.4$), and narrowest
($S\simeq 1$), respectively, for the model discussed here.
}
\label{Fig2}%
\end{figure*}
\newpage
\begin{figure*}
\centering
\includegraphics{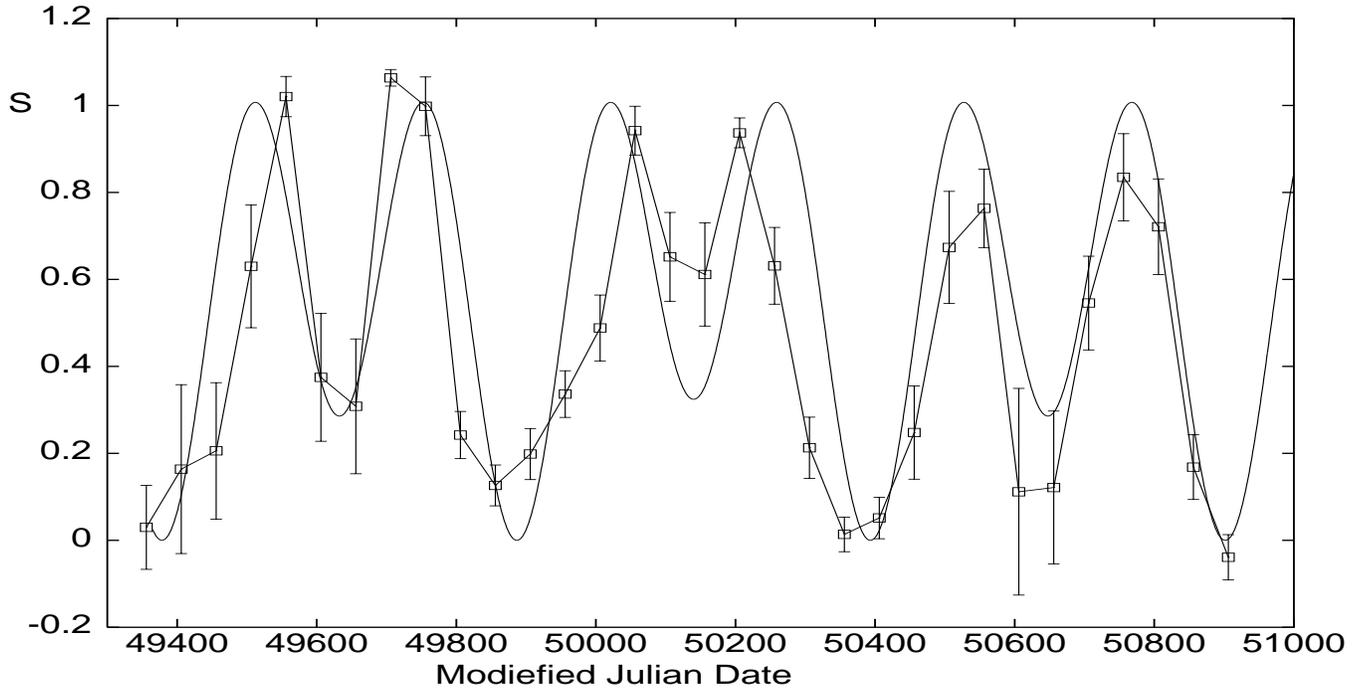}
\caption{ 
The pulse shape parameter data $S$ for PSR B1828-11
(from Stairs et al. \cite{SLS00}).
The shape parameter is defined as
$S=\frac{A_N}{A_N+A_W}$ where $A_N$ and $A_W$ are the fitted heights of
the narrower and wider standard profiles respectively, so that $S\simeq 1$
for the narrowest pulses and $S\simeq 0$ for wider ones.
The data points are obtained by averaging $S$
over multiple bins, and the solid curve is the fit calculated with
$P_{\rm pre}=1016$ d, see text.
}
\label{Fig3}%
\end{figure*}

\end{document}